\documentclass[12pt]{cernart}
\usepackage{times}
\usepackage{graphicx}
\usepackage{cite}
\newcommand{\compass}{{\sc Compass}}

\def\phionly{\ensuremath{\slantright{\Phi}}}

\def\phi1860{\ensuremath{\slantright{\Phi}(1860)}}
\def\phimm1860{\ensuremath{\slantright{\Phi}(1860)^{--}}}
\def\xionly{\ensuremath{\slantright{\Xi}}}
\def\xim   {\ensuremath{\slantright{\Xi\,}^-}}
\def\ximin {\ensuremath{\slantright{\Xi}(1320)^-}}
\def\xip   {\ensuremath{\slantright{\overline{\Xi}\,}^+}}

\def\xi1530{\ensuremath{\slantright{\Xi}(1530)^0}}
\def\xibar1530{\ensuremath{\slantright{\overline{\Xi}}(1530)^0}}

\def\pionly{\ensuremath{\pi}}
\def\pipm  {\ensuremath{\pi^\pm}}
\def\pim   {\ensuremath{\pi^-}}
\def\pip   {\ensuremath{\pi^+}}
\def\thetap{\ensuremath{\slantright{\Theta}^+}}
\def\ldonly{\ensuremath{\slantright{\Lambda}}}
\def\ldbar {\ensuremath{\slantright{\overline{\Lambda}}}}

\newcommand{\slantright}[1]{{#1}}

\begin{document}
\begin{titlepage}
\docnum{CERN--PH--EP/2005--009}
\date{ 5 March 2005}

\vspace{1cm}
\title{\Large Search for the $\phi1860$ Pentaquark at \compass}

\author{\large The COMPASS Collaboration}

\vspace{3cm}
\begin{abstract}
Narrow \xim\pipm\ and \xip\pipm\ resonances produced by quasi-real photons
have been searched for by the COMPASS experiment at CERN.
The study was stimulated by the recent observation of an exotic
baryonic state decaying into \xim\pim ,  at a mass of 1862~MeV,
interpreted as a pentaquark. While the ordinary hyperon states \xi1530 
and \xibar1530 are clearly seen, no exotic baryon is observed in the
data taken in 2002 and 2003.
\vfill
\submitted{(Submitted to The European Physics Journal C)}
\end{abstract}

\newpage
\begin{Authlist}
{\large  The COMPASS Collaboration}\\[\baselineskip]
%
%
E.S.~Ageev\Iref{protvino},
V.Yu.~Alexakhin\Iref{dubna},
Yu.~Alexandrov\Iref{moscowlpi},
G.D.~Alexeev\Iref{dubna},
A.~Amoroso\Iref{turin},
B.~Bade\l ek\Iref{warsaw},
F.~Balestra\Iref{turin},
J.~Ball\Iref{saclay},
G.~Baum\Iref{bielefeld},
Y.~Bedfer\Iref{saclay}
% ...
\end{Authlist}
%
%
\Instfoot{bielefeld}{ Universit\"at Bielefeld, Fakult\"at f\"ur Physik, 33501 Bielefeld, Germany\Aref{d}}
\Instfoot{bochum}{ Universit\"at Bochum, Institut f\"ur Experimentalphysik, 44780 Bochum, Germany\Aref{d}}
\Instfoot{bonniskp}{ Universit\"at Bonn, Helmholtz-Institut f\"ur  Strahlen- und Kernphysik, 53115 Bonn, Germany\Aref{d}}
\Instfoot{bonnpi}{ Universit\"at Bonn, Physikalisches Institut, 53115 Bonn, Germany\Aref{d}}
\Instfoot{brno}{Institute of Scientific Instruments, AS CR, 61264 Brno, Czech Republic\Aref{e}}
\Instfoot{burdwan}{ Burdwan University, Burdwan 713104, India\Aref{g}}
\Instfoot{calcutta}{ Matrivani Institute of Experimental Research \& Education, Calcutta-700 030, India\Aref{h}}
\Instfoot{dubna}{ Joint Institute for Nuclear Research, 141980 Dubna, Moscow region, Russia}
\Instfoot{erlangen}{ Universit\"at Erlangen--N\"urnberg, Physikalisches Institut, 91054 Erlangen, Germany\Aref{d}}
\Instfoot{freiburg}{ Universit\"at Freiburg, Physikalisches Institut, 79104 Freiburg, Germany\Aref{d}}
\Instfoot{cern}{ CERN, 1211 Geneva 23, Switzerland}
\Instfoot{heidelberg}{ Universit\"at Heidelberg, Physikalisches Institut,  69120 Heidelberg, Germany\Aref{d}}
\Instfoot{helsinki}{ Helsinki University of Technology, Low Temperature Laboratory, 02015 HUT, Finland  and University of Helsinki, Helsinki Institute of  Physics, 00014 Helsinki, Finland}
\Instfoot{licerec}{Technical University in Liberec, 46117 Liberec, Czech Republic\Aref{e}}
\Instfoot{lisbon}{ LIP, 1000-149 Lisbon, Portugal\Aref{f}}
\Instfoot{mainz}{ Universit\"at Mainz, Institut f\"ur Kernphysik, 55099 Mainz, Germany\Aref{d}}
\Instfoot{miyazaki}{University of Miyazaki, Miyazaki 889-2192, Japan\Aref{i}}
\Instfoot{moscowlpi}{Lebedev Physical Institute, 119991 Moscow, Russia}
\Instfoot{munichlmu}{Ludwig-Maximilians-Universit\"at M\"unchen, Department f\"ur Physik, 80799 M\"unchen, Germany\Aref{d}}
\Instfoot{munichtu}{Technische Universit\"at M\"unchen, Physik Department, 85748 Garching, Germany\Aref{d}}
\Instfoot{nagoya}{Nagoya University, 464 Nagoya, Japan\Aref{i}}
\Instfoot{praguecu}{Charles University, Faculty of Mathematics and Physics, 18000 Prague, Czech Republic\Aref{e}}
\Instfoot{praguectu}{Czech Technical University in Prague, 16636 Prague, Czech Republic\Aref{e}}
\Instfoot{protvino}{ State Research Center of the Russian Federation, Institute for High Energy Physics, 142281 Protvino, Russia}
\Instfoot{saclay}{ CEA DAPNIA/SPhN Saclay, 91191 Gif-sur-Yvette, France}
\Instfoot{telaviv}{ Tel Aviv University, School of Physics and Astronomy, 
              69978 Tel Aviv, Israel\Aref{j}}
\Instfoot{triestictp}{ ICTP--INFN MLab Laboratory, 34014 Trieste, Italy}
\Instfoot{triest}{ INFN Trieste and University of Trieste, Department of Physics, 34127 Trieste, Italy}
\Instfoot{turin}{ INFN Turin and University of Turin, Physics Department, 10125 Turin, Italy}
\Instfoot{warsaw}{ So{\l}tan Institute for Nuclear Studies and Warsaw University, 00-681 Warsaw, Poland\Aref{k} }
\Instfoot{warsawtu}{ Warsaw University of Technology, Institute of Radioelectronics, 00-665 Warsaw, Poland\Aref{l} }
\Anotfoot{a}{Also at IST, Universidade T\'ecnica de Lisboa, Lisbon, Portugal}
\Anotfoot{b}{Also at University of East Piedmont, 15100 Alessandria, Italy}
\Anotfoot{c}{On leave of absence from JINR Dubna}               
\Anotfoot{d}{Supported by the German Bundesministerium f\"ur Bildung und Forschung}
\Anotfoot{e}{Suppported by Czech Republic MEYS grants ME492 and LA242}
\Anotfoot{f}{Supported by the Portuguese FCT - Funda\c{c}\~ao para
               a Ci\^encia e Tecnologia grants POCTI/FNU/49501/2002 and POCTI/FNU/50192/2003}
\Anotfoot{g}{Supported by UGC-DSA II grants, Govt. of India}
\Anotfoot{h}{Supported by  the Shailabala Biswas Education Trust}
\Anotfoot{i}{Supported by the Ministry of Education, Culture, Sports,
               Science and Technology, Japan}
\Anotfoot{j}{Supported by the Israel Science Foundation, founded by the Israel Academy of Sciences and Humanities}
\Anotfoot{k}{Supported by KBN grant nr 621/E-78/SPUB-M/CERN/P-03/DZ 298 2000 and
               nr 621/E-78/SPB/CERN/P-03/DWM 576/2003--2006 and by MNII research funds for 2005--2007}
\Anotfoot{l}{Supported by  KBN grant nr 134/E-365/SPUB-M/CERN/P-03/DZ299/2000}

\vfill

\hbox to 0pt {~}
\end{titlepage}
%
%
%

\section{Introduction}\label{sec::introduction}

Since many years hadron spectroscopy is concerned with the search
for new hadrons and their subsequent spectroscopy. Particularly
interesting are systems with a flavour content different from the
usual baryonic $qqq$ and $\bar{q}\bar{q}\bar{q}$ or the mesonic
$q\bar{q}$ structure. In the last two years evidence has been
reported on new states interpreted as pentaquarks with a quark
content of $qqqq\bar{q}$. The observation of a state at about
1540~MeV named \thetap~\cite{PDG,web} has been reported by a large
number of experiments but has also been refuted by others,
where the positive observations stem mostly from photoproduction.

Another candidate for an exotic pentaquark state named \phi1860\
(originally called $\Xi$ (1860)) has only been reported by a
single experiment (NA49) using a high-energy proton
beam \cite{na49}. Subsequent \phionly\ searches performed with
different beams and reactions, however, were to no avail
~\cite{aleph,babar,cdf,e690,focus,HERA,hermes,wa89,ZEUS}. Searches
for the \phi1860\  in (possibly advantageous) photoproduction
experiments at a photon virtuality $Q^2 > 1$~GeV$^2$ were performed
by the ZEUS \cite{ZEUS} and HERMES Collaborations \cite{hermes}.
The statistics in these experiments---judged by the number of
observed \xim\ events---was however similar or even below the one
of the NA49 experiment. A high-statistics search using
high-energy real photons ($E_{\gamma} \geq 50$~GeV) was
conducted by the FOCUS Collaboration~\cite{focus}.

The COMPASS experiment at CERN, set up to study the nucleon
structure using high-energy muon scattering, has also collected a
large sample of doubly strange \xim\ baryons in the final state
from quasi-real photoproduction. This paper describes the search
for the doubly strange pentaquark system (and its antiparticle),
produced inclusively in muon-nucleon interactions and decaying
according to
$\phimm1860\rightarrow\xim\pim\rightarrow\discretionary{}{}{}\ldonly\,\pim\pim
      \discretionary{}{}{}\rightarrow \discretionary{}{}{}p\,\pim\pim\pim$.
The data are dominated by quasi-real photoproduction, $Q^2\ll1$~GeV$^2$.
We compare the results to the observation of the well-known state \xi1530\,
which has the same decay chain except for the charge of the pion
in the first decay, $\xi1530\rightarrow\xim\pip$.

\section{The experimental setup}\label{sec::compass}

COMPASS is a fixed-target experiment
at the CERN Super Proton Synchrotron (SPS) using high-energy
muon and hadron beams. The apparatus is described in detail 
in Refs.~\cite{mallot,g1_paper}.
The data for this analysis have been taken in the years 2002 and 2003
with a 160~GeV polarised muon beam hitting a solid-state $^6$LiD target,
polarised longitudinally (80\% of the beam time)
and transversely (20\%) with respect to the beam direction.
The direction of polarisation was flipped regularly, such that any
possible polarisation dependence averages out in this analysis.
The target consists of two cells, each 60~cm long and 3~cm in diameter,
separated by a 10~cm gap.

The beam tracks were reconstructed in a beam telescope consisting of
silicon microstrip and scintillating fibre detectors.
The interaction products are observed in a forward magnetic
spectrometer, with two stages for low and high momenta, respectively,
each equipped with high-resolution tracking detectors and
electromagnetic and hadronic calorimetry.
Information from the RICH detector installed in the first stage of the
spectrometer was not used in this analysis.
Muons were identified downstream of iron and concrete walls.

\begin{figure}[t]
  \begin{center}
    \includegraphics[width=\hsize]{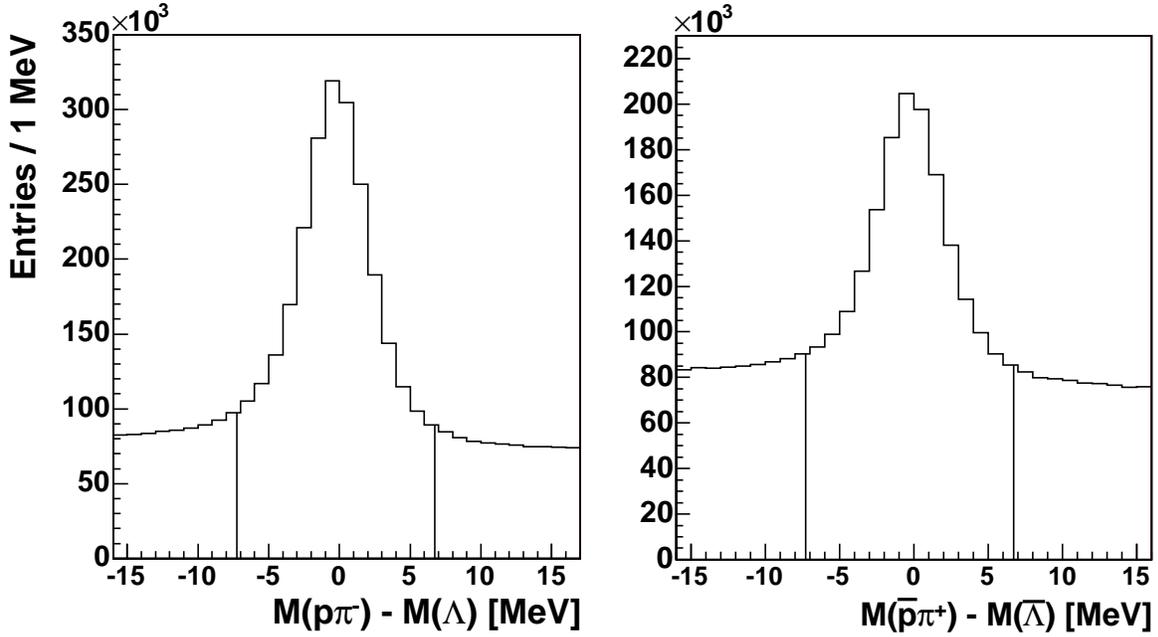}
    \caption{
     The $p\pim$ ($\overline{p} \pi^+$) invariant mass spectra
     with respect to the nominal \ldonly\ mass.
     The vertical lines show the mass cut ($\pm 3\sigma$)
     selecting events for further analysis
     (cf.\ Sect.~\ref{section:Xi-sel}).}\label{Fig::L0}
  \end{center}
\end{figure}

\section{Analysis}\label{sec::analysis}

The initial data set consists of 1.8$\cdot$10$^9$ raw events.
The tracks and momenta of charged particles
were reconstructed in both spectrometer stages.
Using the tracks of the incoming and scattered muon together with
other outgoing charged particle tracks,
the primary interaction point was determined
and was required to be located within the target volume.
Events without a reconstructed primary interaction point
were discarded.
Long-living strange particles were detected by their decay, appearing
as a secondary vertex downstream of the primary interaction point.

\subsection{The \ldonly\ and \ldbar\ selection}
The \ldonly\ decay into $p$ and \pim appears as a secondary
vertex with two outgoing particles of opposite charge.
From their four-momenta
the invariant mass of the \ldonly\ candidate was reconstructed.
In order to reduce the background under the signal, 
the \ldonly\ decay point was required to be located downstream of the target 
volume. Secondary vertices caused by photon conversion,
$\gamma \rightarrow e^+ e^-$, were removed by
requiring $|\cos \theta^*| < 0.9$, with
$\theta^*$ being the c.m.s.\ emission-angle of the negative particle
with respect to the flight direction of the \ldonly.
For the reconstruction of \ldbar, the mass assignment of the two charged
particles was inverted.
The resulting $p\pim$ ($\overline{p}\pip$) invariant mass spectra
are presented in Fig.~\ref{Fig::L0}. They were fitted by the sum of
a Gaussian for the \ldonly\ signal and 
a polynomial parametrising the background.
The total numbers of \ldonly\ (\ldbar) particles
are given in Table~\ref{Tab::Statistics}.

\subsection{The \xim\ and \xip\ selection}\label{section:Xi-sel}
\begin{figure}[t]
  \begin{center}
    \includegraphics[width=\hsize]{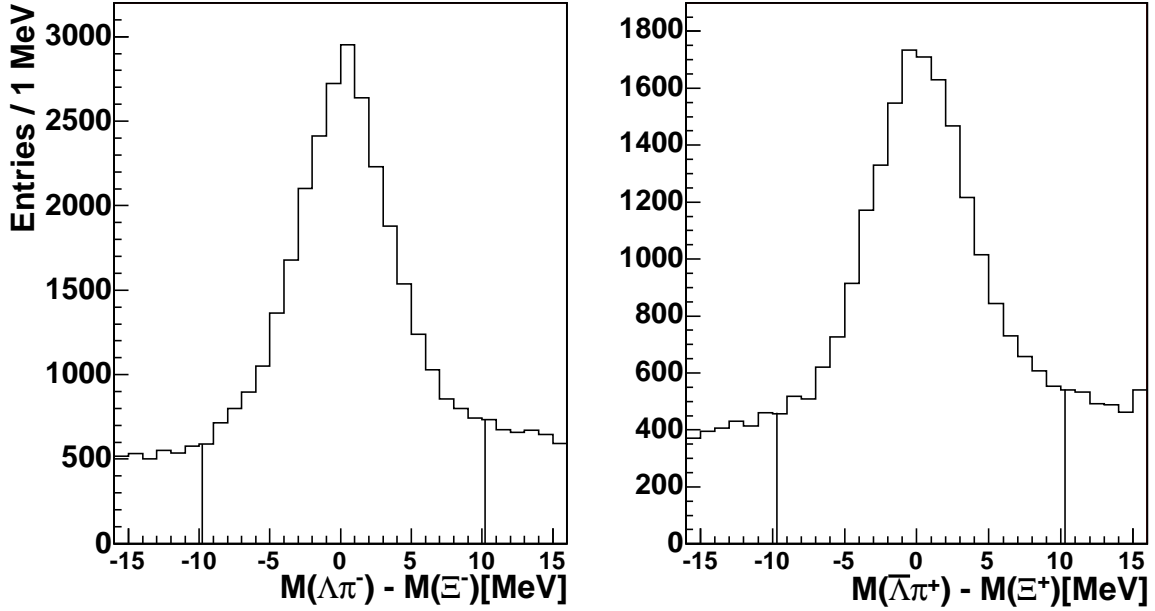}
    \caption{The $\ldonly\pi^-$ ($\ldbar\pi^+$) invariant mass spectra
      relative to the nominal \xim\ mass.
     The vertical lines show the mass cut ($\pm 3\sigma$)
     selecting events for further analysis.
}\label{Fig::Xi}
  \end{center}
\end{figure}

The \ldonly\ candidates in the mass range marked in Fig.~\ref{Fig::L0}
were selected for the further analysis with the nominal \ldonly\ mass
assigned to all  candidates.
The closest distance of approach (CDA) of the
\ldonly\ line of flight to all other tracks,
corresponding to negatively charged
particles and not connected to the primary vertex, was calculated.
For the CDA smaller than 0.8~cm, a possible \xim\ decay point
was reconstructed, 
and only combinations with this decay point downstream of the
primary interaction point and upstream of the \ldonly\ decay point
were retained.
These cuts have been tuned to minimise the background
without reducing the \xim\ signal.
In Figure~\ref{Fig::Xi} the \ldonly\pim\ (\ldbar\pip) invariant mass spectra
with respect to the \xim\ (\xip) nominal mass are shown.
The total numbers of \xim\
and \xip\ particles obtained by the fit procedure described for the
\ldonly (\ldbar) signals are given in Table \ref{Tab::Statistics}.

\subsection{The \xionly\pionly\ mass spectra}

\begin{figure}[tp]
  \begin{center}
    \includegraphics[width=0.9\hsize]{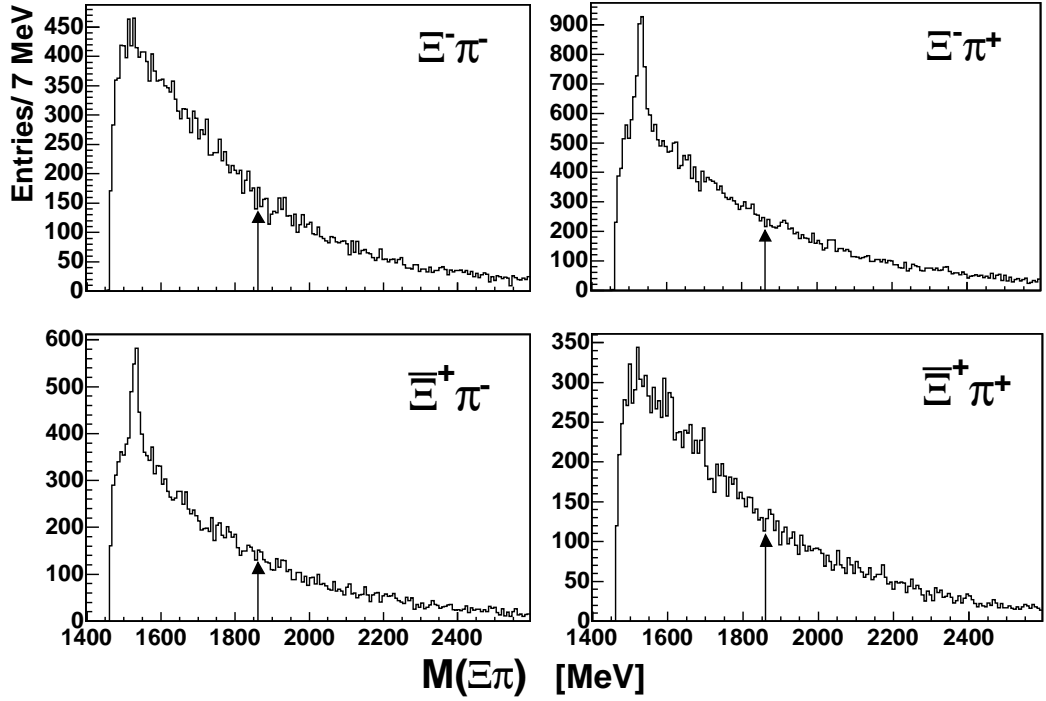}
    \caption{$\Xi\pi$ invariant
             mass spectra of the four possible charge
             combinations. Arrows indicate the position of the
             \phi1860\ signal~\cite{na49}\label{Fig::Pq}.}
  \end{center}
\end{figure}

\begin{figure}[bhpt]
  \begin{center}
    \includegraphics[width=0.9\hsize]{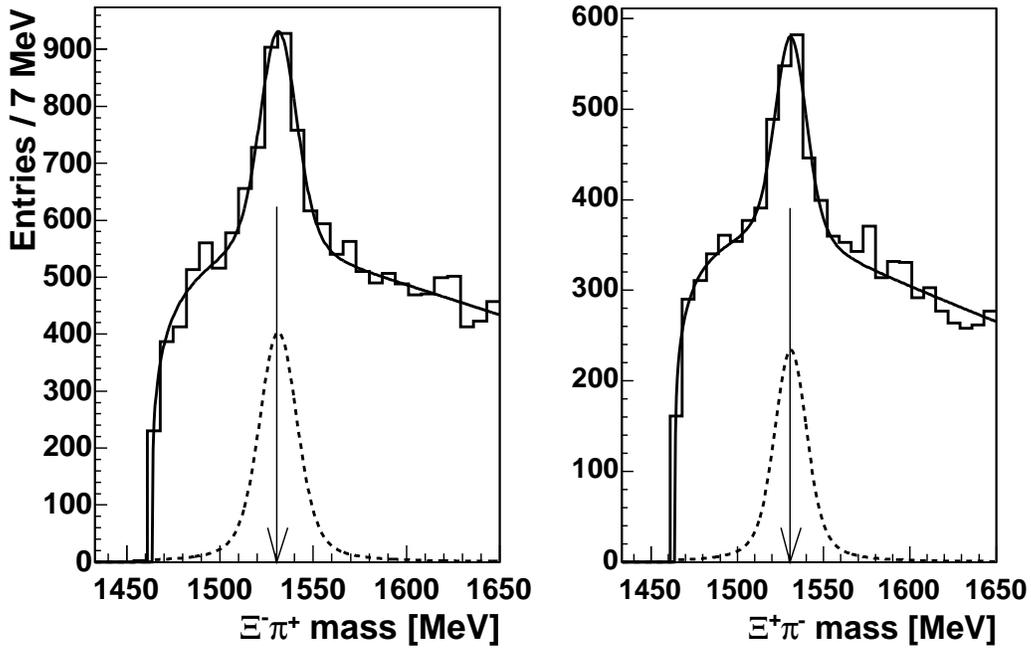}
    \caption{The $\Xi^- \pi^+$ ($\overline{\Xi}^{~+}\pi^-$)
      invariant mass spectra around 1530~MeV.
      Solid lines mark the fit results as described in the text, dashed
      lines mark the fitted signal shape only.
      Arrows indicate the \xi1530 mass.
}\label{Fig::X0fit}
  \end{center}
\end{figure}

The \xi1530 and \xibar1530 resonances are so short-living that their
decay point coincides with the primary interaction point.
All charged particles whose tracks point back to the primary interaction
point, and which were not used in the earlier steps of the reconstruction
or identified as muons, were assumed to be pions. They were combined
with the \xim\ or \xip\ candidates in the mass range
indicated in Fig.~\ref{Fig::Xi}.
The effective masses of the combinations were calculated
assuming the nominal  masses for the  \xim (\xip ).
The mass spectra
for the four charge combinations are shown in Fig.~\ref{Fig::Pq}.
The \xi1530 (\xibar1530) resonance signal is clearly visible.
The invariant mass distributions around their nominal mass are presented
enlarged in Fig.~\ref{Fig::X0fit}. 
The numbers of \xi1530 and \xibar1530
are given in Table~\ref{Tab::Statistics}.
They were  obtained by fits using a Voigtian for the signal 
and a background parametrised as 
$(m-m_0)^{p_1}\cdot\exp(-p_2\cdot m-p_3\cdot m^2)$.

\begin{table}[t]
  \caption{Number of particles as obtained by fitting the mass spectra
    in Figs~\ref{Fig::L0}, \ref{Fig::Xi} and \ref{Fig::X0fit}. 
    The errors take into account the uncertainty in the background
    parametrisation.}
\vspace{2mm}
\small
  \begin{center}
    \begin{tabular}{|c|r@{\,}c@{\,}l|r@{\,}c@{\,}l|}
      \hline
      &\multicolumn{3}{c|}{Yield of particles}
      &\multicolumn{3}{c|}{Yield of antiparticles}   \\
      \hline
     \ldonly / \ldbar \rule{0mm}{4mm}
                 & 1 250 000 & $\pm$ &50 000& \ \ 640 000 & $\pm$ &30 000 \\
      \xim / \xip    &18 000 & $\pm$ &   900&  10 600 & $\pm$ &  600 \\
    \parbox[t]{2cm}{\xi1530 /\\ \raggedleft\xibar1530}       & 1 700 & $\pm$ &   100&     920 & $\pm$ &   80 \\
      \hline
    \end{tabular}
  \end{center}
   \label{Tab::Statistics}
\end{table}

\begin{table*}[t]

\vspace{2mm}

 \caption{\label{tab::pqexp}Summary of \phi1860 searches in
inclusive production. The energies given in column~3  refer to the beam energy 
in case of fixed-target experiments and to $\sqrt{s}$ in case of collider
experiments. The $x_F$ values were calculated assuming a nucleon target.}
 \begin{center}
  \begin{tabular}{|l|c|r@{\,}c@{\,}l|c|r|r|r|}
  \hline
  Experiment&Initial   & \multicolumn{3}{c|}{Energy}     &approx. $x_F$ & \multicolumn{3}{c|}{Yield of} \\
            & state    & \multicolumn{3}{c|}{[GeV]} & for $\Xi^-$   & $\Xi^-\ \ $ & $\Xi$(1530)$^0$ &\phimm1860\\ \hline
   COMPASS& $\mu^+\;\; A$
          & $E_{\mu^+}$&=&160    & $>0$   &  18000  &  1700 &  $<$79    \\

  \hline
   NA49\cite{na49}   & $pp$  
          & $E_p$&=&158   &$-$0.25...0.25   &   1640  &   150 &  36       \\
  \hline
   ALEPH \cite{aleph} & $e^+e^-$   
          &  $\sqrt{s}$&=&$m_{Z^0}$      &     &   3450  &   322 &  $<$24    \\
   BaBar \cite{babar} & $e^+ e^-$ 
          &  $\sqrt{s}$&=&$m_{\Upsilon(4S)}$    &    & 258000  &  17000     & not seen   \\
   CDF   \cite{cdf} & $p \overline p$ 
          &   $\sqrt{s}$&=&1960    &    &  35722  &  2182 & $<$63  \\
   E690  \cite{e690} & $pp$    
          &    $E_p$&=&800           &    & 512850  & 70000 & $<$200 \\
   FOCUS \cite{focus} & $\gamma p$  
          &  $E_{\gamma}$&$\leq$& 300 &    & 800000  & 59391 & $<$170 \\
   HERA-B \cite{HERA} & $p A$    
          &  $E_p$&=&920   &$\approx 0$   &  12000  &  1400 & $<$56  \\
   HERMES \cite{hermes} & $e^- D$
          &  $E_e$&=&27.6  &              & 450&   35 & $<$5      \\
   WA89   \cite{wa89} & $\Sigma^- A$
          & $E_{\Sigma^-}$&=&340 & $>0.1$ & 676000  &  60000   & $<$760  \\
   ZEUS   \cite{ZEUS} &  $e p$
          & $\sqrt{s}$&=&310   &        &   1561  & 192\hbox to 0pt{$^\dag$} & not seen \\
   \hline
  \end{tabular}
\end{center}
$^\dag$sum of \xi1530\ and \xibar1530
\label{tab::experiments}
\end{table*}

The \xi1530\ (\xibar1530) resonance has a natural width 
$\Gamma=9.1$~MeV~\cite{PDG}. The observed widths of the \xi1530\ 
and \xibar1530\ signals in  Fig.~\ref{Fig::X0fit} are  compatible
with the mass resolution of our spectrometer, $\sigma\approx 7$~MeV, expected
from simulation. Assuming a similar width for the \phi1860 signal, we
determined upper limits for the signal of such a state.

The upper limits were obtained as follows: We estimated the expected
background from a polynomial fit to the mass spectra  of the like-sign pairs 
(cf.\ Fig.~\ref{Fig::Pq}), excluding the region from 1825~MeV to 1895~MeV.
We then investigated three intervals of 28~MeV width, 
staggered by 14~MeV, with the central interval centred at 1860~MeV.
The numbers of entries in these intervals are denoted by $n_1, n_2, n_3$.
With the estimated background for each interval, $b_1, b_2, b_3$,
we determined the quantity $\max_{i=1,2,3}(3\sqrt{b_i}+\max(0,n_i-b_i))$ and
thus deduced upper limits for a possible excess of events:
79 events for the \xim\pim\ and 89 events
for the \xip\pip\ final state at a confidence level of 99\%.
Also, no narrow peaks are visible around 1860~MeV
in the spectra for 
the non-exotic unlike-sign pairs.

\section{Discussion and conclusion}\label{sec::conclusions}
This negative result obtained in high-statistics
quasi-real photoproduction is in line with other searches listed in Table~2.

\begin{figure}[t]
  \begin{center}
    \includegraphics[width=0.8\hsize,bb=12 60 527 500]{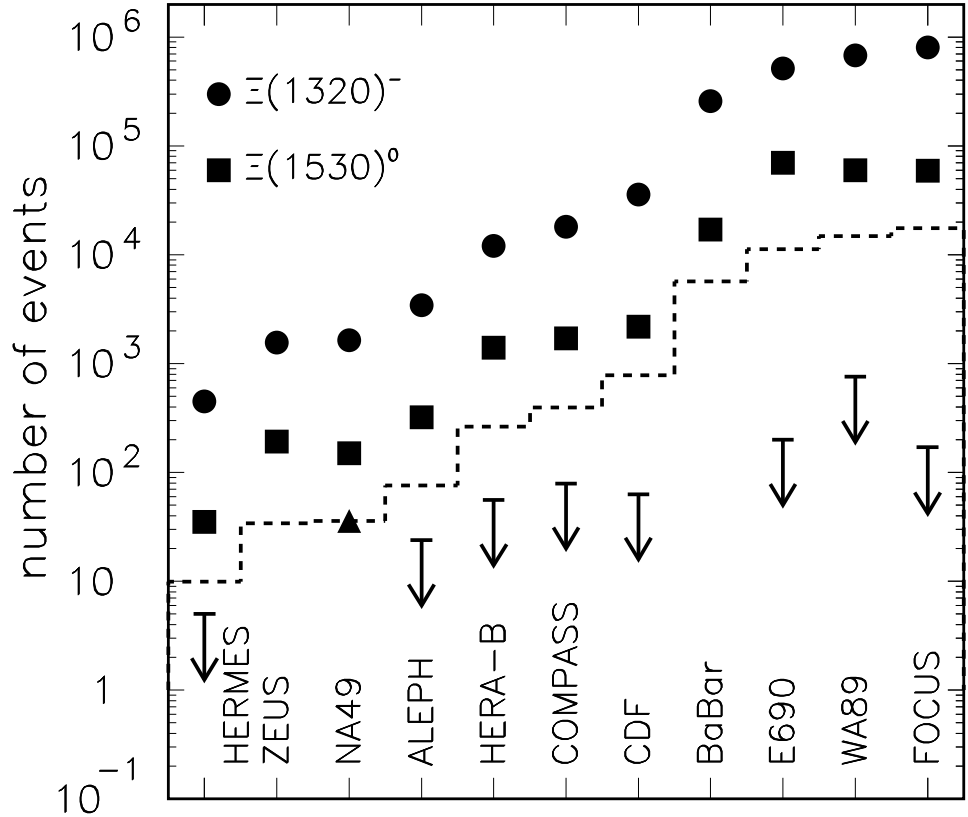}
    \caption{Compilation of the data from
        Table~\ref{tab::experiments}. The dashed line shows the expected
        yield of \phimm1860 from Ref.~\cite{na49} assuming the same
        \phimm1860/\ximin\ event ratio in all experiments.}
  \label{Fig::experiments}
  \end{center}
\end{figure}

A graphical compilation of all search results is shown in 
Fig.~\ref{Fig::experiments}. Note that some data
\cite{babar,cdf,e690,focus,ZEUS} have been shown at conferences only
and are not yet published in journals and therefore should be considered as
preliminary. The circles and squares
in Fig.~5 give the observed numbers of \xim\ and \xi1530\
particles, respectively. The triangle indicates the yield of \phi1860\ observed
by the NA49 Collaboration \cite{na49}.
The arrows show the upper limits reported by the various experiments. Note
that some limits are at 95\% confidence level while others---like the
data presented in this work---are at 99\% confidence level. 
Also, the widths of the examined mass windows vary as the
experimental resolutions are different. Thus a direct comparison
of  the various experimental limits has to consider systematic relative
shifts by a factor of the order of two. The observed
\xim (1320)/\xi1530\ yield ratio varies by at most a factor of
2.5 between the different experiments. 
Assuming  that the relative detection efficiencies for the various
resonances are similar in the various experiments and ignoring possible
differences in the production mechanism of a pentaquark, one can estimate the
expected yields of \phi1860\ pentaquarks by scaling 
the observed numbers of \xim (1320) with the
\phi1860/\xim (1320)  ratio reported by NA49. 
The result is shown by the dashed line
in Fig.~5.
It is obvious that the comparatively large \phimm1860/\discretionary{}{}{}\ximin\ 
production
ratio reported in Ref.~\cite{na49} is not supported by
other experiments. A confirmation of the \phi1860\ existence is thus
still missing.

\section*{Acknowledgements}
Special thanks are due to M.~Zavertiaev for valuable comments and 
discussions during the preparation of this paper. We gratefully acknowledge the support 
of the CERN management and staff
and the skill and effort of the technicians of our collaborating
institutes. 

\begin{center}
\rule{5cm}{0.1mm}
\end{center}


\begin{thebibliography}{00}
\bibitem{PDG} Particle Data Group, S.~Eidelman {\it et al.},
         Phys.\ Lett.\ B {\bf 592} (2004) 1.
\bibitem{web} For recent reviews of experimental results and theoretical
         models see:\\
         http://www.rcnp.osaka-u.ac.jp/$\sim$hyodo/research/Thetapub.html;\\
         A.R.~Dzierba, C.A.~Meyer and A.P.~Szczepaniak,
         {\sl Reviewing the evidence for pentaquarks},
         Proc.\ 1st Meeting of the APS Topical Group on Hadronic Physics, 
         Batavia, Illinois, October 2004 and hep-ex/0412077.
\bibitem{na49} NA49 Collaboration, C.~Alt {\it et al.}, Phys.\ Rev.\ Lett. {\bf 92} (2004) 042003 and priv. comm.
\bibitem{aleph} ALEPH Collaboration, S.~Schael {\it et al.},
         Phys.\ Lett.\ B {\bf 599} (2004) 1.
\bibitem{babar} BaBar Collaboration, B.~Aubert {\it et al.},
         SLAC-PUB-10992 and hep-ex/0502004, submitted to Phys.\ Rev.\ Lett.
\bibitem{cdf}   CDF Collaboration,   O.~Litvintsev {\it et al.},
         hep-ex/0410024~v2, presented at BEACH04, Chicago, USA, July 2004.
\bibitem{e690}  E690 Collaboration, D.~Christian  {\it et al.},
         talk presented at QNP04, Bloomington, USA, May 2004,
         http://www.qnp2004.org.
\bibitem{focus} FOCUS Collaboration, K.~Stenson  {\it et al.},
         hep-ex/0412021, presented at DPF meeting, 
         Riverside, USA, August 2004.
\bibitem{HERA} HERA--B Collaboration, I.~Abt {\it et al.}, 
         Phys.\ Rev.\ Lett.\ {\bf 93} (2004) 212003.
\bibitem{hermes} HERMES Collaboration, A.~Airapetian {\it et al.},
         Phys.\ Rev.\ D {\bf 71} (2005) 032004,
         some values taken from figures therein.
\bibitem{wa89} WA89 Collaboration, M.I.~Adamovich {\it et al.},
         Phys.\ Rev.\ C {\bf 70} (2004) 022201(R) 
         and Eur.\ Phys.\ J.~C. {\bf 11} (1999) 271.
\bibitem{ZEUS}  ZEUS  Collaboration, S.~Chekanov {\it et al.}, DESY 05-018 
         and hep-ex/0501069, submitted to Phys.\ Lett.\ B.
\bibitem{mallot} 
         COMPASS Collaboration, G.K.~Mallot, 
         Nucl.\ Instrum.\ Meth.\ A {\bf 518} (2004) 121.
\bibitem{g1_paper} COMPASS Collaboration, E.S.~Ageev {\it et al.},
         {\sl Measurement of the spin structure of the deuteron 
         in the DIS region}, CERN-PH-EP-2005-001  and   hep-ex/0501073,
         submitted to Phys.\ Lett.\ B.
\end{thebibliography}
\end{document}